\documentclass[letterpaper, 10 pt, conference]{ieeeconf}

\IEEEoverridecommandlockouts %

\usepackage{amsmath,amsfonts,amssymb}
\usepackage{mathtools}
\usepackage{amsfonts} %
\usepackage{mathrsfs} %
\usepackage{bbold} %
\usepackage{pgfplots}
\usepackage[ruled, vlined, linesnumbered]{algorithm2e}
\usepackage{subfig}
\usepackage{tabularx}
\usepackage{multirow}
\usepackage{siunitx}
\usepackage{listings}
\usepackage{flushend}
\usepackage[export]{adjustbox}

\usepackage{caption}
\usepackage{graphicx}
\usepackage{transparent}
\graphicspath{{figures/}}

\usepackage{xargs} %

\usepackage[style=ieee, backend=biber, bibencoding=utf8, maxbibnames=3, mincitenames=1, maxcitenames=2, url=false, doi=false, isbn=false, citestyle=numeric-comp, natbib=true, eprint=false]{biblatex}

\usepackage[colorinlistoftodos,prependcaption,textsize=small]{todonotes}
\graphicspath{ {figures/} }

\newcommandx{\change}[2][1=]{\todo[inline,linecolor=blue,backgroundcolor=blue!25,bordercolor=blue,#1]{#2}}

\usepackage{bm} %
\usepackage{lipsum}
\newcommand{\Secref}[1]{Section~\ref{#1}} 

\newcommand{\TB}[2]{\ensuremath{B^{#1}_{#2}}}
\newcommand{\ASB}[2]{\ensuremath{b^{#1}_{#2}}}
\newcommand{\ASE}[2]{\ensuremath{e^{#1}_{#2}}}

\pgfplotsset{compat=newest}
\pgfplotsset{plot coordinates/math parser=false}
\usetikzlibrary{decorations.markings, positioning}

\newlength\figureheight 
\newlength\figurewidth

\newcommand\copyrighttext{%
	\scriptsize \textcolor{blue}{\textcopyright 2020 IEEE. Personal use of this material is permitted.  Permission from IEEE must be obtained for all other uses, in any current or future media, including reprinting/republishing this material for advertising or promotional purposes, creating new collective works, for resale or redistribution to servers or lists, or reuse of any copyrighted component of this work in other works}}
\newcommand\copyrightnotice{%
	\begin{tikzpicture}[remember picture,overlay]
	\node[anchor=north,yshift=-7.5pt] at (current page.north) {\fbox{\parbox{\dimexpr\textwidth-\fboxsep-\fboxrule\relax}{\copyrighttext}}};
	\end{tikzpicture}%
}

\title{\LARGE \bf BARK: Open Behavior Benchmarking in Multi-Agent Environments}
\author{Julian Bernhard$^{1*}$, Klemens Esterle$^{1*}$, Patrick Hart$^{1*}$, Tobias Kessler$^{1*}$%
	\thanks{$^{1}$fortiss GmbH, Research Institute of the Free State of Bavaria, Munich, Germany}%
	\thanks{$^{*}$These authors contributed equally to this work.}%
}

\DeclareTextFontCommand{\tsf}{\tiny\sffamily} %
\begin{document}

\maketitle
\copyrightnotice
\thispagestyle{empty}
\pagestyle{empty}

\global\csname @topnum\endcsname 0
\global\csname @botnum\endcsname 0

\newcommand{\figurename}{Fig. }

\newcommand {\vect} {\boldsymbol}
\newcommand {\matr} {\boldsymbol}

\newcommand{\state} {\vect{x}}
\newcommand{\stateSpace} {\vect{\mathcal{X}}}
\newcommand{\beliefstate} {\vect{b}}
\newcommand{\contr} {\vect{u}}
\newcommand{\contrSpace} {\vect{\mathcal{U}}}
\newcommand{\meas} {\vect{y}}
\newcommand{\procNoise}{\vect{w}}

\newcommand {\cov}  {\matr{\Sigma}}

\newcommand{\stateNoDelta}{\hat\state}
\newcommand{\contrNoDelta}{\hat\contr}
\newcommand{\procNoiseNoDelta}{\hat\procNoise}

\newcommand{\abc}[2][\empty]{%
  \ifthenelse{\equal{#1}{\empty}}
    {no opt, mand.: \textbf{#2}}
    {opt: \textbf{#1}, mand.: \textbf{#2}}
}

\newcommand {\noiseu} {\procNoise}
\newcommand {\covu} {\matr{\Sigma_{\noiseu,}}}

\newcommand {\noiseuNoDelta} {\procNoiseNoDelta}
\newcommand {\covuNoDelta} {\matr{\Sigma_{\noiseuNoDelta,}}}

\newcommand {\defnoiseu}[1][\empty]{
 \ifthenelse{\equal{#1}{\empty}}
    {\noiseu\sim N(0,\covu)}
    {\noiseu_{#1}\sim N(0,\covu_{#1})}
}

\newcommand {\defnoiseuNoDelta}[1][\empty]{
 \ifthenelse{\equal{#1}{\empty}}
    {\noiseuNoDelta\sim N(0,\covuNoDelta)}
    {\noiseuNoDelta_{#1}\sim N(0,\covuNoDelta_{#1})}
}

\newcommand {\covm} {\matr{R}}
\newcommand {\noisem} {\vect{\nu}}
\newcommand {\defnoisem}[1][\empty]{
 \ifthenelse{\equal{#1}{\empty}}
    {\noisem\sim N(0,\covm)}
    {\noisem_{#1}\sim N(0,\covm_{#1})}
}

\newcommand{\stateB}{\vect{\xi}}
\newcommand{\contrB}{\vect{\nu}}
\newcommand{\procNoiseB}{\vect{\omega}}

\newcommand{\AB}{\mathcal{A}}
\newcommand{\BB}{\mathcal{B}}
\newcommand{\WB}{\mathcal{W}}
\newcommand{\costStateB}{\mathcal{Q}}
\newcommand{\costContrB}{\mathcal{R}}
\newcommand{\covStatesB}{\mathcal{S}_{\state}}
\newcommand{\covProcNoiseB}{\mathcal{S}_{\procNoise}}

\newcommand{\FB}{\mathcal{F}}

\newcommand{\cct}{\vect{t}}
\newcommand{\ccsval}{s}
\newcommand{\ccT}{\matr{T}}
\newcommand{\ccsvec}{\vect{s}}

\newcommand{\costState}{\matr{Q}}
\newcommand{\costContr}{\matr{R}}
\newcommand{\feedbackMatrix}{\matr{K}}
\newcommand{\cost}{J}

\newcommand{\stateConstraintMatrix}{\matr{C}}
\newcommand{\stateConstraintVector}{\vect{c}}

\newcommand{\stateConstraintFunc}{c}

\newcommand{\contrConstraintMatrix}{\matr{D}}
\newcommand{\contrConstraintVector}{\vect{d}}

\newcommand{\contrConstraintFunc}{d}

\newcommand{\stateRef}{\state^{*}}
\newcommand{\contrRef}{\contr^{*}}

\newcommand{\stateDelta}{\Delta\state}
\newcommand{\contrDelta}{\Delta\contr}

\newcommand {\Comment}[1]{\textcolor{blue}{#1}}

\newcommand {\partialder}[4][\bigg]{\frac{\partial #2}{\partial #3}#1|_{#4}}
\newcommand {\partialdernoarg}[3][\bigg]{\frac{\partial #2}{\partial #3}#1}

\newcommand{\nat}{\mathbb{N}}
\newcommand{\real}{\mathbb{R}}
\newcommand{\compl}{\mathbb{C}}

\newcommand{\norm}[1]{\left\| #1 \right\|}

\newcommand{\half}{\frac{1}{2}}

\newcommand{\parenth}[1]{ \left( #1 \right) }
\newcommand{\bracket}[1]{ \left[ #1 \right] }
\newcommand{\accolade}[1]{ \left\{ #1 \right\} }
\newcommand{\pardevS}[2]{ \delta_{#1} f(#2) }
\newcommand{\pardevF}[2]{ \frac{\partial #1}{\partial #2} }

\newcommand{\vecii}[2]{\begin{pmatrix} #1 \\ #2 \end{pmatrix}}
\newcommand{\veciii}[3]{\begin{pmatrix}  #1 \\ #2 \\ #3	\end{pmatrix} }
\newcommand{\veciv}[4]{\begin{pmatrix}  #1 \\ #2 \\ #3 \\ #4	\end{pmatrix}}

\newcommand{\matii}[4]{\left[ \begin{array}[h]{cc} #1 & #2 \\ #3 & #4 \end{array} \right]}
\newcommand{\matiii}[9]{\left[ \begin{array}[h]{ccc} #1 & #2 & #3 \\ #4 & #5 & #6 \\ #7 & #8 & #9	\end{array} \right]}

\newcommand{\transp}{^{\intercal}}
\newcommand{\Reg}{$^{\textregistered}$}
\newcommand{\reg}{$^{\textregistered}$ }
\newcommand{\Tm}{\texttrademark}
\newcommand{\tm}{\texttrademark~}
\newcommand {\bsl} {$\backslash$}

\newtheorem{theorem}{Theorem}[section]
\newtheorem{lemma}[theorem]{Lemma}
\newtheorem{corollary}[theorem]{Corollary}
\newtheorem{remark}[theorem]{Remark}
\newtheorem{definition}[theorem]{Definition}
\newtheorem{equat}[theorem]{Equation}
\newtheorem{example}[theorem]{Example}
\newcommand{\insertfigure}[4]{ %
	\begin{figure}[htbp]
		\begin{center}
			\includegraphics[width=#4\textwidth]{#1}
		\end{center}
		\vspace{-0.4cm}
		\caption{#2}
		\label{#3}
	\end{figure}
}

\newcommand{\refFigure}[1]{\figurename \ref{#1}}
\newcommand{\refChapter}[1]{Chapter \ref{#1}}
\newcommand{\refSection}[1]{Section \ref{#1}}
\newcommand{\refParagraph}[1]{Paragraph \ref{#1}}
\newcommand{\refEquation}[1]{(\ref{#1})}
\newcommand{\refTable}[1]{Table \ref{#1}}
\newcommand{\refAlgorithm}[1]{Algorithm \ref{#1}}

\newcommand{\rigidTransform}[2]
{
	${}^{#2}\!\mathbf{H}_{#1}$
}

\newcommand{\code}[1]
 {\texttt{#1}}

\newcommand{\comment}[1]{\marginpar{\raggedright \noindent \footnotesize {\sl #1} }}

\newcommand{\clearemptydoublepage}{%
  \ifthenelse{\boolean{@twoside}}{\newpage{\pagestyle{empty}\cleardoublepage}}%
  {\clearpage}}

\newcommand{\etAl}{\emph{et al.}\mbox{ }}

\newcommand{\todoi}[1]{\todo[inline]{#1}}

\begin{abstract}
Predicting and planning interactive behaviors in complex traffic situations presents a challenging task. 
Especially in scenarios involving multiple traffic participants that interact densely, autonomous vehicles still struggle to interpret situations and to eventually achieve their own mission goal.
As driving tests are costly and challenging scenarios are hard to find and reproduce, simulation is widely used to develop, test, and benchmark behavior models.
However, most simulations rely on datasets and simplistic behavior models for traffic participants and do not cover the full variety of real-world, interactive human behaviors.
In this work, we introduce BARK, an open-source behavior benchmarking environment designed to mitigate the shortcomings stated above.
In BARK, behavior models are (re-)used for planning, prediction, and simulation.
A range of models is currently available, such as Monte-Carlo Tree Search and Reinforcement Learning-based behavior models.
We use a public dataset and sampling-based scenario generation to show the inter-exchangeability of behavior models in BARK.
We evaluate how well the models used cope with interactions and how robust they are towards exchanging behavior models. Our evaluation shows that BARK provides a suitable framework for a systematic development of behavior models.
\end{abstract}

\IEEEpeerreviewmaketitle

\section{Introduction}
\label{sec:introduction}
Among other technical and legal challenges, a major problem in the development of autonomous vehicles is solving complex tasks in close interaction with humans.
Human drivers have developed strategies to adapt to the behavior of other humans, implicitly embedding potential uncertainty about the behavior of surrounding participants into their own plan.

In our work, we see the behavior \TB{i}{t} of an agent $i$ at time $t$ as \emph{its desired future} sequence of physical states encoding the agent's strategy to reach a short-term goal, e.g changing lane. A behavior may deviate from the executed motion in the environment due to errors in trajectory tracking or environmental influences. Further details are given in \Secref{sec:concept}. 

A longstanding challenge in the development of decision-making for autonomous vehicles is to mimic interactive behavior when interacting with human drivers. Since most behavior generation algorithms do not consider interactions, the number of disengagements in interactive situations, such as urban areas, is especially large~\cite{favaro_autonomous_2017}. We refer to planners addressing this challenge as \textit{interactive behavior generation} algorithms. To model interactivity, planners must employ some kind of prediction model of other agents. Though a variety of modeling methods exist \cite{albrecht_autonomous_2018}, it remains unclear how to guarantee their feasibility in domains like autonomous driving.
Errors in the prediction due to incomplete and faulty models can cause huge problems in real-world driving situations.

This outlines the importance of and need for considering interactions in behavior generation.
It also raises the need for algorithms -- either conventional or learning-based -- to be benchmarked on how well they generalize and cope with inaccurate prediction models.
To tackle these problems, research must be conducted systematically in deterministic environments.
It is generally accepted that vehicle dynamics simulation is a suitable tool for speeding up the development and verification of behavior planning algorithms. 
However, existing frameworks for simulating and benchmarking behavior models rarely provide sophisticated behavior models for other agents.
Microscopic traffic simulators can model traffic flow to simulate groups of vehicles but neglect the interactions between them. 
To simulate interactive behavior, ingredients of both modeling methodologies -- vehicle- and traffic-centered -- are required.
We will elaborate on this in \refSection{sec:related_work}.

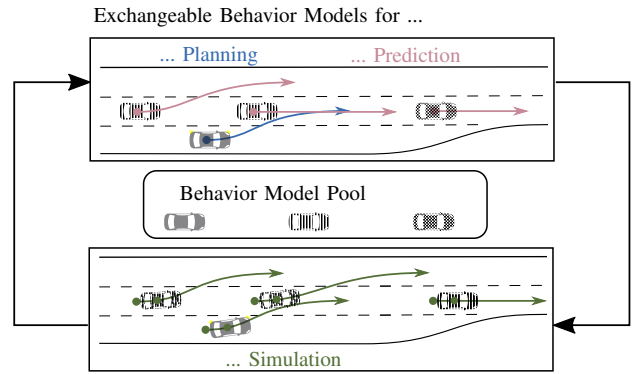
\begin{figure}[t]
	\vspace{0.15cm}
	\footnotesize
	\centering
	\def\svgwidth{0.95\columnwidth}
\begingroup%
  \makeatletter%
  \providecommand\color[2][]{%
    \errmessage{(Inkscape) Color is used for the text in Inkscape, but the package 'color.sty' is not loaded}%
    \renewcommand\color[2][]{}%
  }%
  \providecommand\transparent[1]{%
    \errmessage{(Inkscape) Transparency is used (non-zero) for the text in Inkscape, but the package 'transparent.sty' is not loaded}%
    \renewcommand\transparent[1]{}%
  }%
  \providecommand\rotatebox[2]{#2}%
  \newcommand*\fsize{\dimexpr\f@size pt\relax}%
  \newcommand*\lineheight[1]{\fontsize{\fsize}{#1\fsize}\selectfont}%
  \ifx\svgwidth\undefined%
    \setlength{\unitlength}{3015.83357683bp}%
    \ifx\svgscale\undefined%
      \relax%
    \else%
      \setlength{\unitlength}{\unitlength * \real{\svgscale}}%
    \fi%
  \else%
    \setlength{\unitlength}{\svgwidth}%
  \fi%
  \global\let\svgwidth\undefined%
  \global\let\svgscale\undefined%
  \makeatother%
  \begin{picture}(1,0.59671573)%
    \lineheight{1}%
    \setlength\tabcolsep{0pt}%
    \put(-0.02671387,-0.07993449){\color[rgb]{0,0,0}\makebox(0,0)[lt]{\begin{minipage}{0.00702876\unitlength}\raggedright \end{minipage}}}%
    \put(0.31652826,0.50442798){\color[rgb]{0.21960784,0.42352941,0.69019608}\makebox(0,0)[t]{\lineheight{1.25}\smash{\begin{tabular}[t]{c}... Planning\end{tabular}}}}%
    \put(0.63556134,0.50213145){\color[rgb]{0.77647059,0.5372549,0.59607843}\makebox(0,0)[t]{\lineheight{1.25}\smash{\begin{tabular}[t]{c}... Prediction\end{tabular}}}}%
    \put(0,0){\includegraphics[width=\unitlength,page=1]{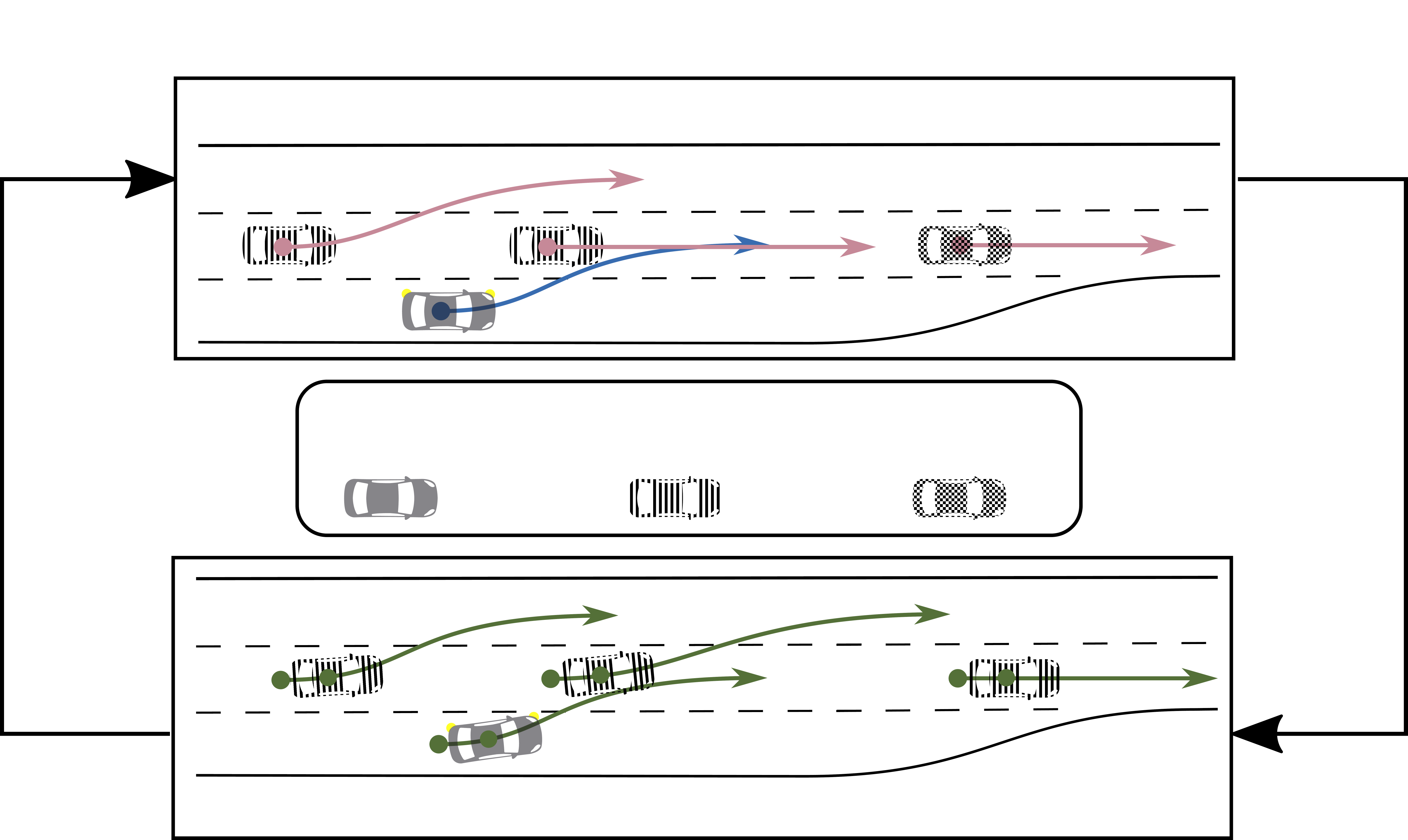}}%
    \put(0.27003554,0.27845247){\color[rgb]{0,0,0}\makebox(0,0)[lt]{\lineheight{1.25}\smash{\begin{tabular}[t]{l}Behavior Model Pool\end{tabular}}}}%
    \put(0.13024903,0.56631281){\color[rgb]{0,0,0}\makebox(0,0)[lt]{\lineheight{1.25}\smash{\begin{tabular}[t]{l}Exchangeable Behavior Models for ...\end{tabular}}}}%
    \put(0.34365836,0.00878815){\color[rgb]{0.32941176,0.43921569,0.21960784}\makebox(0,0)[lt]{\lineheight{1.25}\smash{\begin{tabular}[t]{l}... Simulation\end{tabular}}}}%
  \end{picture}%
\endgroup%

	\caption{The core concept of BARK as a development platform for behavior models: Reuse a behavior model interchangeably for planning, simulation, and prediction.}
	\label{fig:intro}
	\vspace{-0.15cm}
\end{figure}

We claim that a white-box approach for the simulation of agents is the only way to compare multiple approaches on a scientific level.
To systematically improve applied models in these domains, it shall be possible to use a behavior model from a modeling point of view to
\begin{enumerate}
	\item plan the ego motion of the autonomous vehicle,
	\item predict the motion of other, potentially human-driven, vehicles,
	\item simulate an agent in a purely virtual environment.
\end{enumerate}
Furthermore, a platform for behavior model development must support full reproducibility of experiments and provide a fast simulation loop.
This is especially beneficial for machine-learning-based models that require thousands of training episodes. 
\refFigure{fig:intro} sketches the idea of using behavior models interchangeably for planning, prediction, \emph{and} simulation.

In this work, we introduce the simulation platform \textit{BARK}, an acronym for \textit{B}ehavior Benchm\textit{ARK}, which addresses all the needs to develop interactive behavior planners and behavior simulation and prediction models as described in \refSection{sec:concept}.
Besides the tool itself, which is available as open-source software\footnote{\url{https://github.com/bark-simulator/bark/}}, we contribute the methodology to develop and benchmark new behavior models.
We also provide and evaluate reference behavior model implementations based on Reinforcement Learning and Monte-Carlo Tree Search, alongside other simplistic traffic model implementations from the literature. Details on this can be found in \refSection{sec:behavior_models}. 

Besides the contribution to the overall goal of shifting the development efforts from field testing to simulation and modeling, we evaluate the benefits and highlight some of the core features of BARK in \refSection{sec:evaluation}.

\section{Related Work}
\label{sec:related_work}

We will now discuss relevant simulation and benchmarking environments.
Most tools can, to some extent, simulate simple behavior models alongside an accurate motion model of the simulated agents.

CommonRoad\footnote{\url{https://commonroad.in.tum.de/}} was established by \citet{Althoff2017} as a composable benchmarking tool for motion planning on roads. 
It provides a collection of scenarios and benchmarks and aims to evaluate and compare different motion planning algorithms.
With the focus on assessing the performance of motion planners and an easy-to-use interface for controlling one agent in the statically predefined scene, the tool aims to foster comparability between different motion planners. 
CommonRoad is fully open-source.
However, CommonRoad uses only pre-recorded data for the other agents -- only enabling non-interactive behavior planning.
In BARK, the other agents actively interact with the ego vehicle.

The open-source simulator Carla\footnote{\url{http://carla.org/}} has attracted attention recently.
It fills the gap of the research community's need for a high-fidelity simulator with modern software design and state-of-the-art visualization. 
Especially for sensor simulations and visualization of different agent types, the tool provides powerful resources.
\citet{Palanisamy2019} has built a gym environment based on Carla.
However, with Carla being based on the Unreal Game Engine, problems like non-determinism and timing issues are introduced, that we consider undesirable when developing and comparing behavior models.
Contrary to the 1 Hz simulation frequency in Carla, our step environment can cover any type of frequency.
The recently introduced Deepdrive\footnote{\url{https://deepdrive.voyage.auto/}} by Voyage has similar drawbacks as Carla.

A variety of commercial software suites are available.
For an overview of existing automotive simulators, we refer to \citet{Kang2019}.
Most tools accompany specific industry needs and are highly configurable for specific use cases.
However, they are not available as such to the research community and most require significant engineering effort to be adapted to these specific needs.

Microscopic traffic simulators such as SUMO\footnote{\url{https://sumo.dlr.de/docs/}} by \citet{Lopez2018} allow each agent to be simulated individually, and provide a realistic simulation of the traffic flow in a potentially large traffic network with numerous agents.
These tools do not aim for benchmark capacities or tracking the accurate motion of each agent.

\section{BARK in a Nutshell}
\begin{figure}[t]
	\vspace{0.15cm}
	\centering
	\def\svgwidth{8.0cm}
	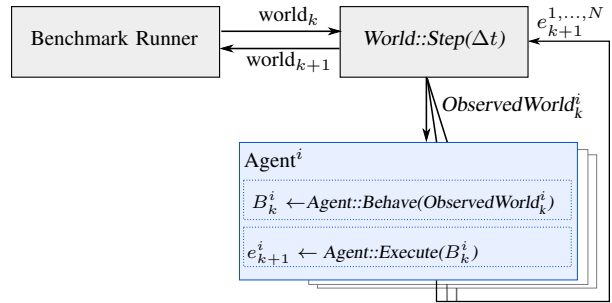
	\caption{BARK's simulation loop is handled by the benchmark runner holding the current world state at discrete world time $k$. In each iteration, the benchmark runner calls \textsl{World::Step($\Delta t$)}.
		This function generates an \textsl{ObservedWorld} for each agent.
		and passes it to the agent's internal \textsl{BehaviorModel} which generates a behavior $\TB{i}{k}$. The behavior is passed to the agent's \textsl{ExecutionModel} calculating the next executed agent state \ASE{i}{k+1}. The next world state at time $t_{k+1} = t_k + \Delta t$ integrates the updated agent states for all agents $N$ and is returned to the \textsl{Benchmark Runner}.}
	\label{fig:runtime}
	\vspace{-0.45cm}
\end{figure}

\label{sec:concept}
BARK focuses on providing a software framework for the systematic evaluation and improvement of behavior models. 
The same implementations can be used to either plan the ego motion of a vehicle, predict other vehicles' motions, and forward simulate a driving scenario.
For example, a traffic model, such as the Intelligent Driver Model \cite{Treiber2000} can on the one hand be used to populate a simulation with agents (cf. \refFigure{fig:intro}) but also as a generative model to predict other agents' motion from the viewpoint of the ego vehicle.
Experiments in BARK shall be fully reproducible, independent of the frequency at which the simulation runs.
To ensure this, BARK models the world as a multi-agent system with agents performing \emph{simultaneous movements} in the simulated world.
At fixed, discrete world time-steps, each agent plans its behavior using an agent-specific behavior model, which only has access to the agent's observed world, but not the simulator's simulated world.
The concept of simultaneous movement ensures that a behavior model can plan based on reproducible input information.
It avoids timing artifacts that may occur in message-passing, middleware-based simulation architectures.
\refFigure{fig:runtime} visualizes the core concept of BARK's simulation model.

\begin{figure}[tb]
	\vspace{0.15cm}
	\footnotesize
	\centering
	\def\svgwidth{\columnwidth}
	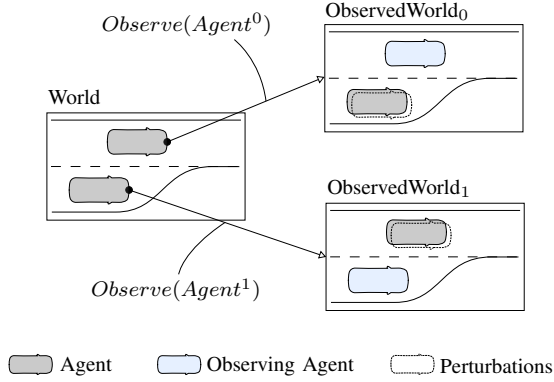
    \caption{Each agent in BARK uses an observed world for defining its behavior.
             In each observed world there is one observing agent (depicted in blue) from whose perspective the observation is being made.
             Perturbations can be introduced by e.g. exchanging the other agents' behavior models and  model parameters in the observed world.}
	\label{fig:observed_world}
	\vspace{-0.15cm}
\end{figure}

BARK uses behavior models not only for behavior planning, but also for predicting other agents in the world.
For example, the observed world of each planner derives from the actual world. 
All agents in this observed world behave according to their prediction configuration.
\refFigure{fig:observed_world} visualizes BARK's observed world model.
With the concept of deriving an observed world with various prediction configurations from the actual world definition, potential errors of the behavior planner caused by inaccurate internal prediction of other traffic participants can be systematically examined. 

BARK provides several state-of-the-art behavior models ranging from conventional planning to machine learning approaches.
BARK's current behavior model implementation will be discussed in \refSection{sec:behavior_models}.
In the following, we take a more detailed look at other BARK components. 

\paragraph{World and ObservedWorld Model} 
\label{sec:observed_world}
The BARK \textsl{World} model contains the map, all objects and agents.
Static and dynamic objects are represented in the form of object lists. We will refer to dynamic objects as agents in the following.

The \textsl{ObservedWorld} model, on the other hand, reflects the world that is perceived by an agent $i$.
BARK's \textsl{ObservedWorld} model accounts for the fact that the observing agent has no access to the true (world) behavior model of other agents.
BARK can model different degrees of observability, by either completely restricting access to the world behavior model or only by perturbing the parameters of the world behavior model. Additionally, occlusions and sensor noise can be introduced in the \textsl{ObservedWorld} model. 

\paragraph{Agent Models}
As shown in  \refFigure{fig:runtime}, an agent in BARK provides two main interfaces: 
\begin{itemize}
 \item \textsl{$\TB{i}{k} \leftarrow $Agent::Behave(ObservedWorld$_\text{k}^{i}$)}: calls the agent-specific behavior model, which generates a behavior trajectory $\TB{i}{k} = (\ASB{i}{1}, \ASB{i}{2}, \ldots, \ASB{i}{L})$, being a sequence of desired future physical agent states $\ASB{i}{l}=(t,x ,y , \theta, v)$ between current simulation world time $t_k$ and at least the end time of the simultaneous movement  $\ASB{i}{L}(t) \geq t_k+\Delta t$. The time discretization of the behavior trajectory can be arbitrary.
 \item \textsl{$\ASE{i}{k+1} \leftarrow$ Agent::Execute(\TB{i}{k})}: calls the agent-specific execution model determining the agent's next state \ASE{i}{k+1} in the world based on the generated behavior trajectory \TB{i}{k}. An additional execution model allows to systematically examine the robustness of behavior planners against execution errors. Since this is not the current focus of our work, we use an interpolation-based execution model. Models that account for controller inaccuracies will form the subject of future work.
\end{itemize}

To implement these two main interfaces, an agent holds the following additional agent-specific information:
\begin{itemize}
  \item \textsl{GoalDefinition}: BARK considers agents to be goal-driven.
        It provides an abstract goal specification with various inherited types of agent goals, e.g. geometric goal regions or lane-based goals.
        Each agent contains a single goal definition instance. 

  \item \textsl{RoadCorridor}: When an agent is initialized, it computes the set of roads and corresponding lanes useful to reach its goal.
  The topology information how roads are connected is extracted from the map.
  Also, the geometric information of the map such as lane boundaries are being discretized. This precomputation avoids computational overhead during simulation.
	\item \textsl{Polygon}: A 2D polygon defines the shape of the agent.
	\end{itemize}

\paragraph{Scenario}
A BARK scenario contains a list of agents with their initial states, behavior and execution models as well as a goal definition for each agent.
Further, it contains a map file in the OpenDrive\footnote{\url{http://www.opendrive.org/}} format.
To support behavior benchmarking, each scenario specifies which agent is considered the `controlled' agent during the simulation.
A BARK scenario does not explicitly specify how agents will behave over time, e.g. using predefined maneuvers or trajectories.
This concept allows us to simulate interactive scenarios in which other agents react to the controlled agent's behavior as well as to data-driven scenarios with fixed agent trajectories.
Both interactive and data-driven scenarios can be combined, as we show in our evaluation.

\paragraph{Scenario Generation}
BARK provides a scenario generation module for configuring sets of scenarios.
The \textsl{ConfigurableScenario} generation used in this work contains source-sink pairs with additional parameter sets.
Using the source and sink, a routing determines a lane on which the agents are placed.
The parameter sets for a source-sink pair specify the distribution of agent states, their behavior and execution models as well as goal definitions.
Using a configuration reader as well as a scenario generation module, the scenario is then built and ready to be executed.
This modularized concept allows us to model a variety of scenario types and enables an easy extension with novel scenario types.

\paragraph{Benchmarking}
For the systematic development of behavior models, BARK supports large scale evaluation of behavior models over a collection of scenario sets contained in a \emph{benchmarking database}.
Binary serialization of the database before experiments are started ensures that scenarios remain reproducible across different systems\footnote{Though BARK supports adjusting random seeds in the scenario generation, differing implementations and versions of pseudorandom number generators across systems may yield differing scenarios using the same random seed.}.

BARK provides a benchmark runner to evaluate specific behavior models with different parameter configurations over the entire benchmarking database. 
The evaluation is based on an abstract evaluator interface calculating a Boolean, integer or real-valued metric based on the current simulation world state. 
The current evaluators available in BARK are:
\begin{itemize}
	\item \textbf{StepCount}: returns the step count the scenario is at.
	\item \textbf{GoalReached}: checks if a controlled agent's \textsl{GoalDefinition} is satisfied.
	\item \textbf{DrivableArea}: checks whether the agent is inside its \textsl{RoadCorridor}.
	\item \textbf{Collision(ControlledAgent)}: checks whether any agent or only the currently controlled agent collided.
	\item \textbf{GoalDistance}: calculates an euclidean distance to the  \textsl{GoalDefinition}.
\end{itemize}
Evaluators can be used not only for benchmarking, but can also internally by the behavior models, e.g.\ for the reward calculation in search or reinforcement learning-based planners. 

The \textit{BenchmarkRunner} runs each scenario of the database evaluating world states of the simulation.
It terminates the scenario run based on criteria defined with respect to the evaluators.
The results are dumped into a \textsc{Pandas} data frame\footnote{\url{https://pandas.pydata.org}}.
BARK also provides a \emph{distributed benchmark runner} based on \textsc{ray}\footnote{\url{https://ray.readthedocs.io}} to perform evaluations in parallel on multiple cores and clusters.
This is especially beneficial for computationally expensive planning algorithms.

\paragraph{Software Design}
BARK has a monolithic, single-threaded core, written in C++.
The core of BARK is wrapped in Python using \textsc{pybind}\footnote{\url{https://github.com/pybind}} including pickling support for all C++ types.
Scenario generation and benchmarking, as well as service methods for parameter handling and visualization, are implemented in Python. 

A simulator planning cycle is entirely deterministic, which enables the simulation and experiments to be reproducible. We argue this to be essential to conduct systematic research in the field of behavior modeling.

The BARK ecosystem is split over multiple Github repositories\footnote{\url{https://github.com/bark-simulator/}}.
We use Bazel\footnote{\url{https://bazel.build/}} as the build system.
Its sandboxed build environment simplifies the reproducibility of experiments since dependency versions can be tracked easily over multiple repositories.

\section{Behavior Models}
\label{sec:behavior_models}

BARK contains a set of reference implementations for behavior models from the literature.
All behavior models inherit from the abstract BehaviorModel class:

\lstset{language=C++,
	basicstyle=\footnotesize\ttfamily,
	keywordstyle=\color{blue}\ttfamily,
	stringstyle=\color{red}\ttfamily,
	commentstyle=\color{green}\ttfamily,
	morecomment=[l][\color{magenta}]{\#}
}  %

\begin{lstlisting}  %

class BehaviorModel {
  public:
    ...
    virtual Trajectory Plan(float delta_time, 
      const ObservedWorld& world)) = 0;
    ...
};
\end{lstlisting}

Behavior models overload the `Plan' function and return a time-dependent state trajectory.
Since all behavior models share the same interface, this makes them easily exchangeable.

By accessing the ObservedWorld, a behavior model can query egocentric semantic information, e.g.\ \textsl{GetLeftLane()$\rightarrow$GetCenterLine()}, as well as semantic relations to other agents, e.g.\ \textsl{GetAgentInFront()}.
Together with the \textsl{Evaluator} concept and implemented standard behavior models, this allows for rapid prototyping and development of new behavior models in BARK.
We list and highlight some of the implemented behavior models in this section.

\paragraph{Intelligent Driver Model} 
The very popular Intelligent Driver Model (IDM) by \citet{Treiber2000} is an easy-to-implement vehicle-following model often used in microscopic traffic simulation.
To model maintaining a gap from the vehicle in front, it leverages the spacing between the agents, speed asymmetries, and realistic braking profiles.
The model cannot handle lane changes, intersections, and unstructured scenarios.
Several extensions of the model exist \cite{Treiber2006, Kesting2010} that consider limitations of human drivers, as well as adaptive cruise control behavior.
To support larger world step times, our IDM implementation assumes that the preceding vehicle maintains a constant velocity throughout a world simulation step.

\paragraph{MOBIL Model}
For multi-lane scenarios, we implemented the MOBIL model, a decision-making entity to trigger lane changes to improve traffic flow \cite{Kesting2015}.
It maintains a politeness factor, which captures how much other agents are slowed down by a potential lane change.
In case of a lane change decision, we start tracking the center line of the respective left or right driving corridor.
The longitudinal motion is controlled using the IDM.

\paragraph{Reinforcement Learning Model}
Reinforcement Learning (RL) learns behaviors implicitly by collecting experiences and updating its policy to maximize the long-term cumulative expected reward.
RL can be divided into three main categories: Actor-only, Critic-only, and Actor-Critic (AC) methods \cite{KondaVijayRandTsitsiklis}.
In particular AC-RL, such as the Proximal-Policy-Optimization (PPO) and the Soft-Actor-Critic (SAC) algorithm, have been shown empirically to work well \cite{Schulman, Haarnoja2018} for continuous control problems.
In this work, we use the SAC algorithm that uses a stochastic policy $\pi(\underline{s}) = \underline{a}$ with the state $\underline{s}$ and action $\underline{a}$.
The SAC agent acts and collects experiences in BARK and improves its policy iteratively.
The other agents may have arbitrary behavior models, such as the IDM model or also use an RL model.
We use `Observers' that transform the semantic BARK environment into a representation suitable for RL.
The reward $r$ is calculated using `Evaluators'.
These modules are available in our Machine Learning module\footnote{\url{https://github.com/bark-simulator/bark-ml}}.
As it integrates the standard OpenAI Gym-interface\footnote{\url{https://github.com/openai/gym}}, various popular RL libaries, such as TF-Agents\footnote{\url{https://github.com/tensorflow/agents}} can be easily integrated used with BARK.

\paragraph{Multi-Agent Monte Carlo Tree Search Model}
Monte Carlo Tree Search (MCTS) is a well-known tree search method for finding approximate solutions to sequential decision problems via sampling \cite{browne_survey_2012}.
It has been adapted to interactive driving by using information sets assuming simultaneous, multi-agent movements of traffic participants \cite{Lenz2016, paxton_combining_2017}.
They apply it to the context of cooperative planning, meaning that they introduce a cooperative cost function, which minimizes the costs for all agents.
Similar to \cite{Lenz2016}, our action set consists of lane changing and lane keeping (CV, CA, CD\footnote{Constant Velocity, Constant Acceleration, Constant Deceleration}).
The MCTS algorithm iteratively performs selection, expansion, rollout, and backpropagation in each search iteration.
As a selection strategy, we are currently using Upper Confidence Trees (UCT).
After expanding a new node, we obtain an instance of ObservedWorld.
During rollout, we randomly select actions for all interacting agents.

\paragraph{Single-Agent Monte Carlo Tree Search Model}
Similar to the Multi-Agent MCTS implementation, we implement a Single-Agent MCTS, in which the other agents are assumed to follow some kind of modeled behavior.
During expansion and rollout, we forward-simulate the observed world using a behavior model for each agent.
These are set via the prediction setup of the observed world before the MCTS algorithm is started.
Any behavior model from BARK with varying parameter specifications can be used.
During expansion, the tree policy is used.
During rollout, we randomly select actions for the ego agent.

\paragraph{Dataset Tracking Model}
To get real-world data into the simulation, BARK contains an agent model that tracks recorded trajectories as close as possible. 
We can start the simulation at an arbitrary timestamp and include all or only a subset of recorded agents.
We evaluate this using the INTERACTION dataset \cite{Zhan2019}.

\section{Evaluation}
\label{sec:evaluation}
We demonstrate the capabilities of BARK in two use cases:
\begin{enumerate}
	\item Employing BARK \textbf{behavior models for behavior prediction} using the sampling-based scenario generation: We benchmark the effect of an inaccurate prediction model on the performance of an MCTS and RL-based planner.
	\item Employing BARK \textbf{behavior planners for behavior simulation} using dataset-based scenario generation: We benchmark how planners perform when replacing human drivers in recorded traffic scenarios.
\end{enumerate} 
Our evaluation demonstrates the advantages of BARK for development, systematic improvement and large scale benchmarking of behavior models, simulation, and prediction algorithms.

\subsection{Behavior Models for Behavior Prediction}

\begin{figure}[t]
	\vspace{0.15cm}
	\centering
	\rotatebox{-90}{	\input{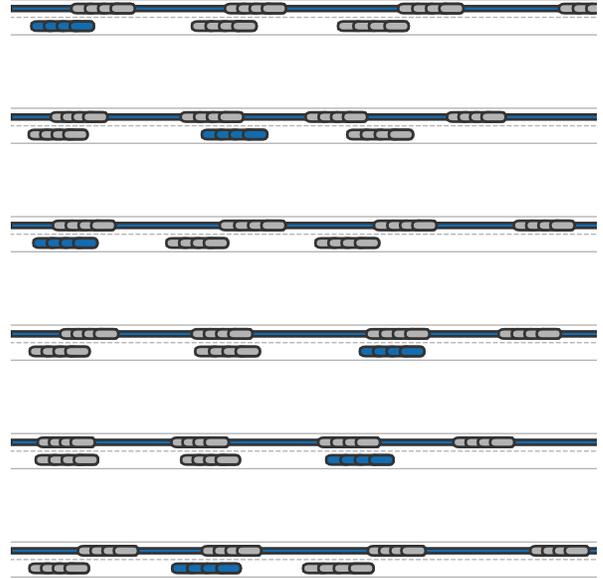}  }
	\caption{The world states of the lane change task after the first four time steps ($t=0.8\,\text{s}$, $\Delta t=0.2 \text{s}$). Scenarios are generated with BARK's sampling-based scenario generation.
           The benchmarked agent (blue) must change to the left lane (goal definition is depicted in light blue). Past agent states are shown with increasing transparency.
           The driving direction is from the left to the right.}
	\label{fig:highway_scenario} 
	\vspace{-0.15cm}
\end{figure}
\label{subsec:evaluation_prediction}
The performance of online planning algorithms, such as the MCTS, depends greatly on the accuracy of the prediction models used internally.
RL can be considered as an offline planning algorithm -- not relying on a prediction model but requiring a training environment to learn an optimal policy beforehand. 
The inaccuracy of prediction relates to the amount of behavior model inaccuracy between training and evaluation. 
In the following, we show to evaluate both sources of errors - inaccurate prediction models and training data - under a common framework.

In our experiment, we consider lane changing in crowded urban roads as depicted in \refFigure{fig:highway_scenario}. 
We evaluate the performance of MCTS and RL planners statistically, over a large set of scenarios.
The scenarios are generated as outlined in \refSection{sec:concept}. We employ uniform sampling of longitudinal distances from the range $[20\,\text{m}, 30\,\text{m}]$ and velocities from the range $[40\,\text{km/h}, 60\,\text{km/h}]$.
All vehicles except the ego vehicle are controlled using the IDM. 
In the benchmark database, the following configurations are stored:
\begin{itemize}
  \item 0\% variation: Vehicles in this scenario use the same IDM model parameters as employed by the MCTS for behavior prediction and RL for training.
        The IDM parameter time headway, accounting for a velocity-dependent safety distance, is set to $T_\text{Head} = 3s$.
	\item 20\%, 40\%, and 80\% variation: In three additional scenario sets, we decrease $T_\text{Head}$ by the given percentages, respectively, e.g. $T_\text{Head}^\text{40\%} = 1.8s$. 
\end{itemize}

In the experiment, BARK's parallelized benchmark runner runs 600 scenarios for each scenario set.
The following planning configurations are applied to the ego vehicle:
\begin{itemize}
	\item Single-Agent MCTS for 2k, 4k, and 8k search iterations
	\item SAC policy trained with RL over scenario set 0\% variation until results were stable (120k episodes, 6h)
\end{itemize}

\begin{figure}[b]
	\vspace{-0.15cm}
	\centering
	\input{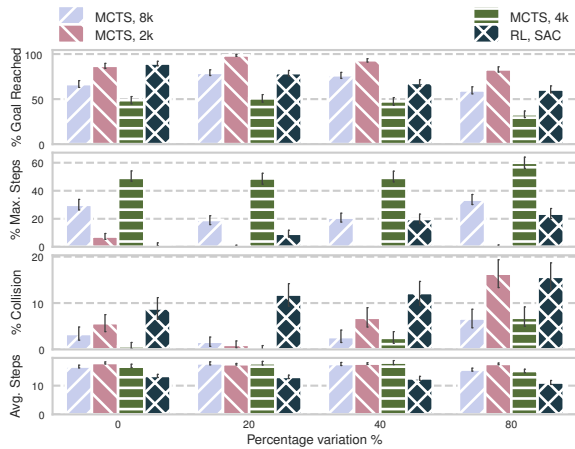}  
	\caption{Statistical benchmark of the MCTS and RL planners for increasing percentages of prediction error.}
	\label{fig:benchmark_results}
	\vspace{0.15cm}
\end{figure}
\refFigure{fig:benchmark_results} gives the percentages of each scenario run with the controlled agent reaching the goal, colliding, or exceeding the maximum allowed simulation time steps ($>$30), as well as the average number of simulation steps in the event that the goal is reached. 
It is clear that the collision rate rises as the prediction error increases.
For 2k iterations, the MCTS fails to explore worst-case outcomes explaining the high success but also high collision rate. 
For 4k iterations, the worst-case outcomes are explored, yet no goal-reaching plan is found, yielding low collision but a large maximum number of steps. 
With a higher exploration at 8k iterations, the success rate increases greatly, with a slight increase in collisions.
The RL, SAC model outperforms the MCTS model for 0\% variation in terms of the success rate.
However, when the other agent's behavior is different from that used in training, the collision rate rises more quickly.

Our experiment showed that a \emph{systematic} evaluation of behavior planners, modeling limited knowledge about other agents' behavior is necessary to uncover and understand the weaknesses of both conventional and learning-based approaches in such settings.
BARK covers the whole development life cycle to tackle this open problem.

\subsection{Behavior Planners for Behavior Simulation}
\label{subsec:evaluation_simulation}
Planning models should be evaluated using realistic real-world data. 
However, when a planner is inserted into recorded scenarios, others will keep the behavior as specified in the dataset, yielding an open-loop simulation.
Since this restricts the evaluation of interactive scenarios, we require behavior models to adapt to other dataset-based agent models. 
We evaluate the suitability of BARK's existing behavior models for this use case by inserting them into a recorded traffic scenario without fine-tuning the model parameters in any way.

\begin{figure}[t]
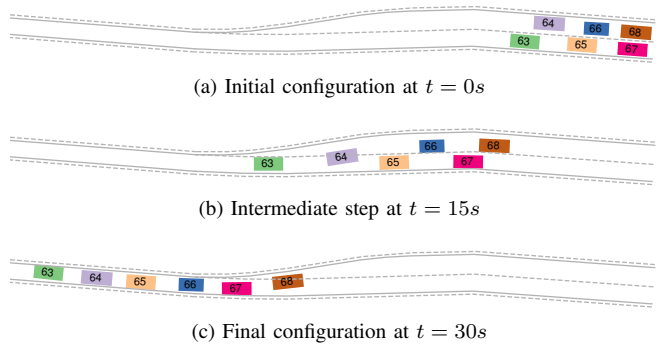

	\vspace{0.15cm}
   	\centering
	\subfloat[Initial configuration at $t=0s$\label{subfig-1:dummy}]{%
		\input{figures/interaction_dataset_0.pgf}
	}
	\hfill
	\subfloat[Intermediate step at $t=15s$\label{subfig-2:dummy}]{%
		\input{figures/interaction_dataset_149.pgf}
	}
	\hfill
	\subfloat[Final configuration at $t=30s$\label{subfig-3:dummy}]{%
		\input{figures/interaction_dataset_299.pgf}
	}
	\caption{Replay of the INTERACTION dataset at a simulation step size $\Delta t=0.1s$.}
	\label{fig:scenario_mer_deu}
	\vspace{-0.15cm}
\end{figure}

Datasets in the prediction and planning community usually consist of fused object lists. 
In contrast to current datasets \cite{Alexiadis2004, Krajewski2018}, the INTERACTION dataset \cite{Zhan2019} also provides maps, which are essential for most on-road planning approaches. 
Here, we analyze the two-way merge scenario {\small{\textsf{DR\_DEU\_Merging\_MT}}}\footnote{The scenario starts at $232.000s$}. 
\refFigure{fig:scenario_mer_deu} shows the original scenario at the initial, the intermediate, and the final time-step. 
The scenario originates in Germany, where vehicles must follow the zip-merge principle. 
The scenario is particularly challenging as the agents not only have to master a merge, but also maintain the order given by the zip-merge rule. Otherwise collisions may occur.

In \refTable{tab:dataset_replay}, we evaluate whether a collision occurred when an agent from the dataset was replaced with various agent behavior models. 
We denote the set of replaced agents by $\mathcal{A}_\square$.
In the evaluation, we do not account for all possible combinations of agents for replacement, but focus on six challenging ones.
We establish that, with the proposed tooling, critical scenarios can be analyzed easily.
Algorithm fine-tuning is considered a subsequent step. 
We compare the performance of the IDM, MOBIL, and Single Agent MCTS described in \refSection{sec:behavior_models}. 
All three models use IDM parameters, as the longitudinal behavior of the MOBIL model is controlled using IDM, while Monte Carlo Tree Search uses an IDM as a prediction model of the other agents. 
We thus define four different parameter sets $\mathcal{P}_{1-4}$ for the IDM, see \refTable{tab:idm_parameters}. 
As we aim to provide a methodology for evaluation, we only vary the minimum distance, without fine-tuning of other model parameters for this single scenario. 
The IDM is a pure lane-following model and thus not capable to perform lane changes.
We thus do not replace agent sets $\mathcal{A}_{0}, \mathcal{A}_{1}, \mathcal{A}_{2}, \mathcal{A}_{6}$ with the IDM.

\begin{table}[b]
\vspace{-0.15cm}
\caption{Agents from the dataset are replaced by a behavior model.
         $\Diamond$ indicates no collision occurred.
         $\blacklozenge$ indicates a collision occurred.
         $\mathcal{A}$ denotes the set of agents we replace.
         $\mathcal{P}$ refers to the respective IDM parameter set being used.}
\centering
	\begingroup
	\setlength{\tabcolsep}{2.3pt} %
	\begin{tabular}{ll|llll|llll|llll}
	 & & \multicolumn{4}{c}{IDM} & \multicolumn{4}{c}{MOBIL} & \multicolumn{4}{c}{MCTS}\\
	$\mathcal{A}$ & Replaced IDs & $\mathcal{P}_1$ & $\mathcal{P}_2$ & $\mathcal{P}_3$ & $\mathcal{P}_4$ & $\mathcal{P}_1$ & $\mathcal{P}_2$ & $\mathcal{P}_3$ & $\mathcal{P}_4$ & $\mathcal{P}_1$ & $\mathcal{P}_2$ & $\mathcal{P}_3$ & $\mathcal{P}_4$ \\ 
	\hline 
	$\mathcal{A}_0$ & 66 & - & - & - & - & $\blacklozenge$ & $\blacklozenge$ & $\Diamond$ & $\Diamond$  & $\blacklozenge$ & $\blacklozenge$  & $\blacklozenge$ & $\blacklozenge$ \\ 
	$\mathcal{A}_1$ & 68 & - & - & - & - & $\blacklozenge$ & $\blacklozenge$  & $\blacklozenge$ & $\blacklozenge$  & $\Diamond$ & $\Diamond$  & $\Diamond$ & $\Diamond$ \\ 
	 $\mathcal{A}_2$ & 66, 68 & - & - & - & - & $\blacklozenge$ & $\blacklozenge$  & $\blacklozenge$ & $\blacklozenge$  & $\blacklozenge$ & $\blacklozenge$  & $\blacklozenge$ & $\blacklozenge$ \\ 
	$\mathcal{A}_3$ & 65 & $\blacklozenge$  & $\blacklozenge$  & $\blacklozenge$ & $\blacklozenge$ & $\blacklozenge$ & $\blacklozenge$  & $\blacklozenge$ & $\blacklozenge$  & $\blacklozenge$ & $\blacklozenge$  & $\blacklozenge$ & $\blacklozenge$ \\
	$\mathcal{A}_4$ & 67 & $\blacklozenge$  & $\blacklozenge$ & $\Diamond$ & $\Diamond$ & $\blacklozenge$ & $\blacklozenge$  & $\Diamond$ & $\Diamond$  & $\blacklozenge$ & $\blacklozenge$  & $\blacklozenge$ & $\blacklozenge$ \\
	$\mathcal{A}_5$ & 65, 67 & $\blacklozenge$  & $\blacklozenge$  & $\blacklozenge$  & $\blacklozenge$ & $\blacklozenge$ & $\blacklozenge$  & $\blacklozenge$ & $\blacklozenge$  & $\blacklozenge$ & $\blacklozenge$  & $\blacklozenge$ & $\blacklozenge$ \\ 
	$\mathcal{A}_6$& 66, 68, 65, 67 & - & - & - & - & $\blacklozenge$ & $\blacklozenge$  & $\blacklozenge$ & $\blacklozenge$  & $\blacklozenge$ & $\blacklozenge$  & $\blacklozenge$ & $\blacklozenge$ \\ 
\end{tabular} 
\endgroup
\label{tab:dataset_replay}
\vspace{0.15cm}
\end{table}

\begin{table}[t]
	\vspace{0.15cm}
	\caption{Parameters of the IDM}
	\centering
	\begin{tabular}{lll|llll}
	Parameter & & & $\mathcal{P}_1$ & $\mathcal{P}_2$ & $\mathcal{P}_3$ & $\mathcal{P}_4$ \\ 
	\hline 
	Desired velocity & $v_0$ & $[m/s]$ & 5 & 5 & 5 & 5\\
	Maximum accel. & $a_{max}$ & $[m/s^2]$ & 1.7 & 1.7 & 1.7 & 1.7\\
	Time gap & $\tau$ & $[s]$ & 1 & 1 & 1 & 1\\
	Comfortable decel. & $b$ & $[m/s^2]$ & 1.7 & 1.7& 1.7 & 1.7\\
	Minimum distance & $s$ & $[m]$ & 2 & 3 & 4 & 5\\
	\end{tabular} 
	\label{tab:idm_parameters}
	\vspace{-0.15cm}
\end{table}

We observe that, without fine-tuning model parameters, most behavior models fail to coexist next to replayed agents.
The IDM can partially solve for $\mathcal{A}_4$, as it needs to retain sufficient gap distance from the preceding vehicle to let merging vehicles in.
This is crucial, as the order must be retained to prevent crashes further on in the scenario.
MOBIL can partially solve $\mathcal{A}_0$, where the agent has to change to the left lane.
If the gap distance parameter is too small, the agent will try to accelerate and change lanes before agent 65.
However, even if the lane change is executed successfully, a collision occurs at the end of the scenario, due to the incorrect order of the vehicles.
This shows the necessity to develop behavior models for simulation that are capable of handling more complex constraints, such as maintaining a specific order. 

Although some MOBIL configurations can master $\mathcal{A}_4$ without collision, the overall scenario is changed, as sometimes lane changes are triggered to go from left to right.
Collision metrics often do not account for the changed scenario.
A wide range of research has been carried out in the field of prediction \cite{Zhan2018} that quantifies the measure of similarities between recorded and modeled behavior. 
We will address this in future work.
All prediction parameter sets used in the MCTS successfully master the lane change of $\mathcal{A}_{1}$. 
Note that we do not constrain the final order, therefore replacing agents in the left lane (65, 67) fails.

We conclude that current rule-based models (IDM, MOBIL) perform poorly in highly dense scenarios, as they do not model obstacle avoidance based on prediction or future interaction. 
More elaborate models should be investigated, as these allow constraints such as ordering and traffic rules to be modeled.
MCTS can be used, but without an accurate model of the prediction, it also leads to crashes, similar to the evaluations in \refSection{subsec:evaluation_prediction}.

\section{Next Steps}
\label{sec:next_steps}
Besides further development of the platform and fostering a community for the systematic improvement of behavior models, we consider the following topics to be the most relevant.

\subsection{Behavior Modeling}
Integrating a more comprehensive set of agent models, e.g.\ based on the formalization of traffic rules \cite{Esterle2019a} as planner and evaluator would be a big benefit for the platform.
Modern optimization-based methods, e.g.\ mixed-integer programming \cite{Kessler2019}, have proven to be computationally fast for motion and strategic planning, and can act as a dual approach compared to learning-based approaches.
A combination of classical and learning-based methods \cite{Hart2019} is computationally fast and achieves safe and comfortable motions.

Recently imitation learning has shown great benefits for agent modeling \cite{Bansal2019, Behbahani2019}. 
Such behavior models promise to have great benefits, especially when used in the simulation.
Robustness against inaccuracies in prediction should be further evaluated, e.g.\ by advancing risk-aware planning \cite{Bernhard2019}.

\subsection{Agent Types}
As of now, only vehicle models are implemented in BARK.
However, other agent types can easily be integrated into BARK due to is highly modular and abstract architecture.
Having other types of agents, such as cyclists or pedestrians would provide additional benefits.
For pedestrians, a variety of datasets and prediction algorithms already exist \cite{Ridel2018}.

\subsection{Scenario Languages}
\label{sec:scenario_todos}
We did not choose to base the scenario definition on a standard like OpenSCENARIO\footnote{\url{http://www.openscenario.org/}}, as the introduced complexity is not relevant for our use case. 
Alternative approaches, such as the work by \citet{Damm2018} modeling scenarios as sequence charts, do not account for behavior model aspects. 
However, having interchangeable scenario definitions between different simulation tools would be highly beneficial.

\subsection{Co-Simulation with Other Environments}
BARK is not designed to replace any existing simulation tool, but its features can enhance other environments and vice versa. 
We aim to publish an interface with Carla to benefit from its high-fidelity visualization and contribute sophisticated agent behavior.
Also supporting state-of-the-art robotics and automotive middlewares would be beneficial to ease the exchange of components.

\section{Conclusion and Future Work}
\label{sec:conclusion}

In this work, we introduced the behavior benchmarking and simulation tool BARK, an environment tailored for developing interactive behavior models.
Behavior models in BARK are exchangeable, which allows them to be used for planning, prediction, and simulation.

In our evaluation, we analyzed how robust the Monte-Carlo Tree Search and Reinforcement Learning behavior models are against prediction inaccuracy.
With BARK's data-set tracking model, we evaluated whether sophisticated behavior planners, as well as simplistic traffic models, can drive accurately together with humans in recorded scenarios.
Both experiments demonstrated the usefulness of having interchangeable behavior models in BARK.

However, the evaluation revealed that more systematic efforts must be made to understand the circumstances under which state-of-the-art behavior models can robustly accomplish prediction, simulation, and planning tasks.
BARK is a platform that enables these efforts.

BARK is open-source under the MIT license and can be applied in a wide range of applications and scenarios.
We welcome contributions from other researchers in this field.

\section*{Acknowledgment}
This research was funded by the Bavarian Ministry of Economic Affairs, Regional Development and Energy, project Dependable AI and supported by the Autonomous Intelligent Driving GmbH.

\renewcommand{\bibfont}{\small}
\printbibliography

\end{document}